\documentclass[a4paper]{article}

\usepackage{INTERSPEECH2020}
\usepackage{amsmath}
\usepackage{amsfonts}%
\usepackage{caption}%
\usepackage{graphicx}
\usepackage{subcaption}
\usepackage{booktabs}
\usepackage{multirow}
\usepackage{placeins}
\usepackage{url}
\usepackage{tikz}
\usepackage[nolist,nohyperlinks]{acronym}
\usepackage{amssymb}
\usepackage{pifont}
\usepackage[nolist,nohyperlinks]{acronym}
\usepackage{multirow}
\usepackage[T1]{fontenc}
\usepackage{xcolor}
\usepackage{url}

\acrodef{WADASNR}{Waveform Amplitude Distribution Analysis}
\acrodef{ETDNN}{Extended TDNN}
\acrodef{CycleGAN}{Cycle Consistent Generative Adversarial Network}
\acrodef{VC}{VoxCeleb concatenated}
\acrodef{SV}{speaker verification}
\acrodef{UEN}{Unsupervised Enhancement Network}
\acrodef{SEN}{Supervised Enhancement Network}
\acrodef{EN}{enhancement network}
\acrodef{BL}{baseline}
\acrodef{UE}{unsupervised enhancement}
\acrodef{SE}{supervised enhancement}
\acrodef{DA}{domain adaptation}
\acrodef{unsup}[\textit{unsup}]{\textit{unsupervised}}
\acrodef{sup}[\textit{sup}]{\textit{supervised}}
\acrodef{adv}[\textit{adv}]{\textit{adversarial}}
\acrodef{cyc}[\textit{cyc}]{\textit{cycle-consistency}}
\acrodef{fmap}[\textit{fmap}]{\textit{feature mapping}}
\acrodef{enh}[\textit{enh}]{\textit{enhancement}}
\acrodef{aug}[\textit{aug}]{\textit{augmentation}}
\acrodef{SNR}{Signal-to-Noise Ratio}
\acrodef{MUSAN}{Music, Speech and Noise}
\acrodef{AMI}{AMI Meeting Corpus}
\acrodef{SITW}{Speakers In The Wild}
\acrodef{RIR}{Room Impulse Response}
\acrodef{SRI}{Stanford Research Institute}
\acrodef{dB}{decibel}
\acrodef{melFB}{mel filter-bank}
\acrodef{aug}[\textit{aug}]{augmentation}
\acrodef{SVS}{Speaker Verification System}
\acrodef{SOTA}{state-of-the-art}
\acrodef{PLDA}{Probabilistic Linear Discriminant Analysis}
\acrodef{VAD}{Voice Activity Detection}
\acrodef{DCT}{Discrete Cosine Transform}
\acrodef{RT}{Reverberation Time}
\acrodef{disc}{discriminator}
\acrodef{LSGAN}{Least Squares Generative Adversarial Network}
\acrodef{minDCF}{minimum Detection Cost Function}
\acrodef{DAN}{Domain Adaptation Network}
\acrodef{conv}[\textit{conv}]{convolutional}
\acrodef{wrt}{with respect to}
\acrodef{WPE}{Weighted Prediction Error}
\acrodef{ASR}{Automatic Speech Recognition}
\acrodef{DFL}{Deep Feature Loss}
\acrodef{DNN}{Deep Neural Network}
\acrodef{FM}{feature mapping}
\acrodef{GAN}{Generative Adversarial Network}
\acrodef{ENs}{Enhancement Networks}

\def\Xmat{\mathbf{X}}
\def\xvec{\mathbf{x}}

\def\tel{\mathrm{tel}}
\def\mic{\mathrm{mic}}
\def\reverb{\mathrm{reverb}}
\def\clean{\mathrm{clean}}
\def\adv{\mathrm{adv}}

\def\FM{\mathrm{FM}}
\def\SEN{\mathrm{SEN}}

\title{Single Channel Far Field Feature Enhancement \\
        For Speaker Verification In The Wild}
\name{Phani Sankar Nidadavolu$^{1,2}$,
Saurabh Kataria$^1$,
Paola Garc\'ia-Perera$^{1,2}$,\\
Jes\'us Villalba$^{1,2}$,
Najim Dehak$^{1,2}$}
\address{
$^1$Center for Language and Speech Processing, $^2$Human Language Technology Center of Excellence,\\Johns Hopkins University, Baltimore, MD, USA}
\email{\{snidada1,skatari1\}@jhu.edu,leibny@gmail.com,\{jvillal7,ndehak3\}@jhu.edu}

\begin{document}

\maketitle
\begin{abstract}
We investigated an enhancement and a domain adaptation approach to make speaker verification systems robust to perturbations of far-field speech.
In the enhancement approach, using \textit{paired} (parallel) reverberant-clean speech, we trained a supervised \ac{GAN} along with a feature mapping loss.
For the domain adaptation approach, we trained a \ac{CycleGAN}, which maps features from far-field domain to the speaker embedding training domain.
This was trained on \textit{unpaired} data in an unsupervised manner.
Both networks, termed \ac{SEN} and \ac{DAN} respectively, were trained with multi-task objectives in (filter-bank) feature domain.
On a simulated test setup, we first note the benefit of using \ac{FM} loss along with adversarial loss in \ac{SEN}.
Then, we tested both supervised and unsupervised approaches on several real noisy datasets.
We observed relative improvements ranging from 2\% to 31\% in terms of DCF.
Using three training schemes, we also establish the effectiveness of the novel \ac{DAN} approach.

\end{abstract}
\noindent\textbf{Index Terms}: Speaker verification, dereverberation, speech enhancement, adversarial feature learning 

\section{Introduction}
\label{sec:introduction}

One of the standard speech enhancement approaches is to learn a mapping function from acoustic features of degraded speech to clean speech using a \ac{DNN}~\cite{xu2014regression}.
This feature mapping (MP) solves the \textit{non-linear regression} problem by minimizing distance metrics like $L_1$ or $L_2$ between the output and reference clean features.
This is a supervised approach, since the \ac{DNN} is trained on \textit{paired} clean-degraded speech, usually obtained by simulation.
The $L_2$ objective trains a regressor that outputs the mean of all plausible outputs, which is known to produce smooth and/or distorted features.~\cite{bishop2006pattern,ledig2017photo}.
This issue is well-noted in enhancement community~\cite{wang2019bridging}.
In this work, we explored the usage of \ac{adv} loss~\cite{goodfellow2014generative} to overcome the distortions introduced by the \ac{FM} approach.
We focused on dereverberation aimed at improving \ac{SVS} performance.

Recently, task specific enhancement approaches are gaining attention in speech research.
The usage of \ac{cyc} loss~\cite{zhou2016learning,zhu2017unpaired} and \ac{adv} loss~\cite{goodfellow2014generative} along with \ac{FM} loss to train denoising networks, for improving  \ac{ASR}, is explored by~\cite{meng2018cycle} and~\cite{meng2018adversarial} respectively.
Results are reported on simulated test conditions.
For feature denoising in \ac{SV}, \ac{DFL}~\cite{germain2018speech} in lieu of \ac{FM} is proposed in~\cite{kataria2020feature,kataria2020analysis}.
Speech enhancement is one of the main approaches considered in developing robust \ac{SVS}s to adverse environments during JSALT 2019 workshop~\cite{garcia2019speaker,nidadavolu2020unsupervised,kataria2020feature,sun2020progressive}.
Also, a relevant non-task specific speech enhancement network using \ac{adv} loss is proposed in~\cite{pascual2017segan}. 

In this work, we explore single channel far-field (reverberant) microphone feature enhancement, in log \ac{melFB} space, for improving \ac{SV}.
Previously, we proposed an \ac{UEN}~\cite{nidadavolu2020unsupervised,nidadavolu2019lr} trained on \textit{unpaired} reverb-clean data, that transforms features from reverberant to clean domain.
\ac{UEN} has shown good dereverberation and denoising capabilities by improving performance on simulated reverberant, simulated noisy, and real datasets collected in wild/uncontrolled environments.
\ac{UEN} also obtained better verification performance compared to the widespread \ac{WPE}-based speech dereverberation approach~\cite{nakatani2010speech,yoshioka2012generalization}.
\ac{UEN}, since does not require \textit{paired} data, can be trained on real data from reverberant (target) and clean (source) domains\footnote{We used the terms clean/source and reverberant/target interchangeably in this paper}, thus avoiding the need for using simulated data.

The main contributions of this work are as follows.
First, we propose a \ac{SEN}, trained on \textit{paired} far-field(reverberant)-clean data, using a multi-task objective--a combination of \ac{FM} and \ac{adv} losses.
Second, we demonstrate the importance of training the \ac{SEN} with \ac{adv} loss by performing an ablation study on the loss functions; and by testing it on \ac{SVS} trained without data augmentation.
Third, we propose a \ac{DAN} that maps features from reverberant to some chosen domain (not necessarily clean).
\ac{DAN}, like \ac{UEN}, is also trained on \textit{unpaired} training data.
Fourth, we test the effectiveness of using \ac{SEN} and \ac{DAN} in improving the performance of a \ac{SVS} trained with data augmentation and three testing schemes.

Our experimental approach was as follows: we developed a \ac{SVS} pipeline where the features of evaluation data (\emph{enrollment} and \emph{test} utterances) were mapped to clean or any chosen domain \textit{via} \ac{SEN} and \ac{DAN} respectively.
We compared them with the previously proposed homogeneous \ac{UEN}-\ac{SVS} pipeline~\cite{nidadavolu2020unsupervised}--\ac{SVS} trained and evaluated on features enhanced using \ac{UEN}.
To make an ideal comparison between \ac{SEN}, \ac{DAN} and \ac{UEN}, we trained all the networks on common list of audio files, unless specified otherwise.

\section{Enhancement Networks}
\label{sec:enh_procedure}

\subsection{Supervised Enhancement Network (SEN)}
\label{ssec:sup_enhancement}

The \ac{SEN} is trained using \textit{paired} reverberant-clean speech to minimize a combination of \ac{FM} and \ac{adv} losses.
We chose $L_1$ metric for the \ac{FM} objective:
\begin{align}
    \label{fmap_obj}
    &L_{\FM}(\SEN, \Xmat_{\reverb}, \Xmat_{\clean}) = \\
    & \mathop{\mathbb{E}_{(\xvec_{\reverb}, \xvec_{\clean}) \sim (p_{\reverb},p_{\clean})}{||\SEN(\xvec_{\reverb})-\xvec_{\clean}||}_1} \nonumber\;.
\end{align}
This objective usually distorts the output by making it smooth~\cite{bishop2006pattern,ledig2017photo}. The added \ac{adv} loss~\cite{goodfellow2014generative} avoids this. \ac{adv} loss requires a discriminator--a binary classifier that discriminates between the enhanced and original clean features. The \ac{SEN} is then trained to trick the discriminator in believing that the output features are sampled from the original clean feature distribution instead of the enhanced feature distribution. At the end of the training, the enhanced and original clean features become indistinguishable by the discriminator, making the enhanced features more realistic, thus avoiding distortion. We used \textit{least-squares} objective~\cite{mao2017least} to train the discriminator as,
\begin{align}
    \label{disc_obj}
    &L_{\mathrm{Disc}}(\SEN, D_{\clean}, \Xmat_{\reverb}, \Xmat_{\clean}) = \nonumber\;\\
   &\mathop{\mathbb{E}_{\xvec \sim p_{\clean}}[(D_{\clean}(\xvec)-1)^2]} \nonumber\;\\
    &+ \mathop{\mathbb{E}_{\xvec \sim p_{\reverb}}[(D_{\clean}(\SEN(\xvec)))^2]}.
\end{align}
The \ac{adv} objective for the \ac{SEN} is
\begin{align} 
    \label{adv_obj}
    &L_{\adv}(\SEN, D_{\clean}, \Xmat_{\reverb}) =\nonumber\;\\ 
    & \mathop{\mathbb{E}_{\xvec \sim p_{\reverb}}[(D_{\clean}(\SEN(\xvec))-1)^2]}.
\end{align}
The final multi-task objective for training the \ac{SEN} is given by   
\begin{align} 
    \label{sen_obj}
    L(\SEN,  D_{\clean}) &= {\lambda}_{\FM} L_{\FM}(\SEN, \Xmat_{\reverb}, \Xmat_{\clean}) \nonumber\;\\
    \quad & + {\lambda}_{\adv} L_{\adv}(\SEN, D_{\clean}, \Xmat_{\reverb})\;,
\end{align}
where ${\lambda}_{\FM}$ and ${\lambda}_{\adv}$ represent the weights assigned to \ac{FM} and \ac{adv} objectives respectively.

\subsection{Unsupervised Enhancement Network (UEN)}
\label{ssec:unsup_enhancement}

We compare \ac{SEN} with the previous \ac{UEN} work~\cite{nidadavolu2020unsupervised}.
\ac{UEN} is trained on \textit{unpaired} data. The procedure is as follows. 
We train a \ac{CycleGAN}~\cite{zhu2017unpaired}, which consists of two generators and two discriminators.
One generator maps features from clean to reverberant domain, while the second maps features from reverberant to clean domain.
Generators are trained using a multi-task objective, consisting of \ac{adv} loss and \ac{cyc} loss.
Similar to \ac{SEN}, \ac{adv} loss is responsible for making the generator to produce features that appear to be drawn from the real distribution of the domain we are mapping to.
The \ac{adv} loss of each generator is obtained using its respective discriminator. The discriminator used least-squares objective (LS-GAN)~\cite{mao2017least}.
The \ac{cyc} loss additionally constrains the generator to reconstruct original features of each domain from the generated features in the opposite domain (achieved by minimizing the $L_1$ distance between original and reconstructed features).
The \ac{cyc} loss makes sure no information is lost during the mapping.  

During evaluation, features of reverberant speech are enhanced by the reverberant-to-clean  generator, termed as \ac{UEN}.
More details on the training procedure and objectives used for training \ac{CycleGAN} can be found in related previous works~\cite{nidadavolu2020unsupervised,nidadavolu2019lr,nidadavolu2019cycle}.

\subsection{Domain Adaptation Network (DAN)}
\label{ssec:dan}

\ac{UEN}, described above, transforms reverberant features to clean domain. Hence, we call the transformation \emph{reverberant feature enhancement} or \emph{feature dereverberation}. On the other hand, \ac{DAN} transforms features from reverberant domain to any chosen domain (details in Section~\ref{ssec:dataset_details}). This mapping is also attained by training a \ac{CycleGAN}, similar to the one trained for \ac{UEN}, except that the source domain does not need to be \textit{clean}.
The \ac{CycleGAN} for \ac{DAN} is trained on \textit{unpaired} data from a selected source domain and reverberant domain. During evaluation, the reverberant features are transferred to the source domain using the corresponding generator in the \ac{CycleGAN}, termed as \ac{DAN}. Except for the difference in training data used, the procedure for training the \ac{DAN} is identical to \ac{UEN}.

\section{Experimental Procedure}
\label{sec:exp_description}

\subsection{Dataset Details}
\label{ssec:dataset_details}

The training of \ac{SEN} required access to \textit{paired} data from clean and reverberant domains, which was obtained as follows. The audio files from the same YouTube video of VoxCeleb1~\cite{nagrani2017voxceleb} and Voxceleb2~\cite{chung2018voxceleb2} were concatenated, denoted as \ac{VC}, to obtain longer audio sequences. Since \ac{VC} was collected in wild conditions and contained unwanted background noise, we filtered the files  based on their \ac{SNR}, estimated by \ac{WADASNR} algorithm~\cite{kim2008robust,nidadavolu2020unsupervised,zen2019libritts,nidadavolu2019lr}. We retained the \ac{VC} files with \ac{SNR}$>19$ \ac{dB}. The high SNR signals, thus obtained, termed as \ac{VC} \textit{clean}, consisted of 1665 hours of speech from 7104 speakers. The far-field data
was obtained \textit{via} simulation by first convolving \ac{VC} \textit{clean} with simulated RIRs\footnote{RIRs available  at \url{http://www.openslr.org/26}} with RT60 values in the range 0.0-1.0 seconds.
Then, assorted \textit{noise} files from \ac{MUSAN}~\cite{snyder2015musan} corpus were artificially added as \textit{foreground} noise (at \ac{SNR} levels of 15, 10, 5, and 0\ac{dB}) to the simulated reverberant speech (\textit{speech} and \textit{music} portions from \ac{MUSAN} were not used in the simulation).
Simulated reverb-clean parallel corpora, termed as \ac{VC} \textit{reverb\_noise}-\ac{VC} \textit{clean}, was used as training data for the \ac{SEN}.  

\begin{table}
    \centering
    \resizebox{\columnwidth}{!}{%
    \begin{tabular}{@{}cccccccc@{}}
        \hline
        \textbf{Network} & \textbf{Output} & \textbf{Training Data} & \textbf{Data Type} & \textbf{Approach} & \textbf{Objectives} \\
        \hline
        \ac{SEN} & \multirow{2}{*}{\textit{clean}} & \ac{VC} \textit{reverb\_noise} \& \ac{VC} \textit{clean} & \textit{paired} & Regression & \ac{FM} \& \ac{adv} \\
        \ac{UEN} & & \ac{VC} \textit{reverb\_noise} \& \ac{VC} \textit{clean} & \multirow{2}{*}{\textit{unpaired}} & \multirow{2}{*}{\ac{CycleGAN}} & \multirow{2}{*}{\ac{cyc} \& \ac{adv}}\\
        \ac{DAN} & \textit{noise} & \ac{VC} \textit{reverb\_noise} \& \ac{VC} \textit{noise} & & &  \\
        \hline
    \end{tabular}%
    }
    \caption{Overview of three enhancement networks}
    \label{tab:networks_summary}
    \vspace{-5mm}
\end{table}

To make a fair comparison between \ac{SEN} and \ac{UEN}, the latter was also trained on \ac{VC} \textit{reverb\_noise}-\ac{VC} \textit{clean}. However, the \textit{unpaired} reverb-clean pairs required for training the \ac{UEN} were drawn randomly without any correspondence between them. The source domain for training the \ac{DAN} was obtained by adding assorted \textit{noise} files from \ac{MUSAN} to \ac{VC} \textit{clean}, termed as \ac{VC} \textit{noise}, while the target domain data (\ac{VC} \textit{reverb\_noise}) remains the same as \ac{UEN}. We added assorted \textit{noise} to the source domain, to increase data variability and obtain better generalization. The choice of the type of noise was based on the observation from the following experiment. We obtained three copies of \ac{VC} by adding \textit{music}, \textit{speech} and \textit{noise} files from MUSAN. We then trained three individual \ac{SVS}s on each of these conditions and observed that system trained on VC \textit{noise} gave best performance compared to systems trained on \textit{music}, \textit{speech} and also \ac{VC} clean on all the evaluation sets considered in this work. Table~\ref{tab:networks_summary} summarizes the training details of three networks.


Once trained, \ac{SEN} and \ac{DAN} networks were used to enhance/adapt the features of \textit{evaluation} corpora, which were finally tested with an x-vector~\cite{snyder2018x} based \ac{SVS}.

We experimented with \ac{SVS}s trained without and with data augmentation.
\ac{SVS} without data augmentation were trained on \ac{VC} \textit{clean}.
For \ac{SVS} trained with augmentation (similar to \cite{snyder2018x}), we used simulated \ac{VC} \textit{reverb\_noise} (described above) and \ac{VC} \textit{additive} as far-field and additive noise corpora to augment \ac{VC} \textit{clean}.
\ac{VC} \textit{additive} corpora consisted of \ac{VC} \textit{noise}, \ac{VC} \textit{babble} and \ac{VC} \textit{music}, each obtained by adding \textit{noise}, \textit{speech} and \textit{music} from \ac{MUSAN} to \ac{VC} \textit{clean} at \ac{SNR}s randomly chosen from 15, 10, 5 and 0 \ac{dB}.
A randomly chosen subset of \ac{VC} \textit{reverb\_noise} and \ac{VC} \textit{additive}, twice the size of \ac{VC} \textit{clean}, was used as augmented data for training the x-vector network.

We tested the \ac{SVS}s on both simulated and real datasets. Simulated reverberant test set was obtained from \ac{SITW}, labelled as \ac{SITW} \textit{reverb}. We treated SITW as clean corpus. SITW \textit{reverb} was created similar to the \ac{VC} \textit{reverb\_noise} except that the maximum value of RT60 for the RIRs used was set to 4.0 seconds (instead of 1.0). We ensured \ac{RIR}s used for training and testing simulations were disjoint.
         
For the real testing conditions, we used three different corpora~\cite{garcia2019speaker} collected in different scenarios: 
\begin{itemize}
\small
    \item \textbf{Meeting} ({\ac{AMI}} \cite{mccowan2005ami}): with a setting of 3 different meeting rooms with 4 individual headset Microphones, 8 Multiple Distant Microphones forming a microphone array; 180 speakers x 3.5 sessions per speaker (sps). Since we are exploring enhancement with single microphone, we focused only on the mix Headset.
    \item \textbf{Indoor controlled} ({\ac{SRI} data \cite{SRI-Real-Voices}}): with a setting of 23 different microphones placed throughout 4 different rooms; controlled backgrounds, 30 speakers x 2 sessions and 40 h,  live speech along with background noises (TV, radio). 
    \item \textbf{Wild} ({\it \textbf{BabyTrain}}): with an uncontrolled setting, 450 recurrent speakers, up to 40 sps (longitudinal), 225hrs; suitable for diarization and detection.
\end{itemize}

The enrollments for verification were generated by accumulating non-overlapping speech (5, 15 and 30s duration) of every target speaker along one or multiple utterances. For the test, we cut the audio into 60 second chunks. We did a Cartesian product between the enrollments and the test segments to generate all possible trials.
Then, based on certain criteria, some trials were filtered out. For example, same session and same microphones were not allowed to produce a target-trial pair.
Table~\ref{tab:datasets_description} shows a summary of all datasets used in this work. 

\begin{table*}[tbp]
    \centering
    \begin{tabular}{@{}c|cccc|cccc|c@{}}
        \hline
        \multicolumn{5}{c|} {\textbf{Train}} &
        \multicolumn{5}{c} {\textbf{Evaluation}} \\ 
        \hline
        \textbf{Real} & \multicolumn{4}{c|}{\textbf{Simulated}} & \multicolumn{4}{c|}{\textbf{Real}} & \textbf{Simulated} \\
        \hline
        \multirow{2}{*}{\ac{VC} \textit{clean}} & \multicolumn{3}{c}{\ac{VC} \textit{additive}} & \multirow{2}{*}{\ac{VC} \textit{reverb\_noise}} & \multirow{2}{*}{\ac{AMI}} & \multirow{2}{*}{\ac{SRI}} & \multirow{2}{*}{BabyTrain} & \multirow{2}{*}{\ac{SITW}} & \multirow{2}{*}{\ac{SITW} \textit{reverb}} \\
        \cmidrule(lr){2-4}
        & \ac{VC} \textit{noise} & \ac{VC} \textit{babble} & \ac{VC} \textit{music} &  &  &  &  &  &   \\ 
        \hline               
    \end{tabular}
    \caption {Summary of the datasets used in training and testing the \ac{SVS}. \ac{VC} stands for VoxCeleb concatenated}
    \label{tab:datasets_description}
    \vspace{-6mm}
\end{table*}

\subsection{Network Architectures}
\label{network_archs}

\ac{SEN} was a fully \ac{conv} residual network with an encoder-decoder architecture. The encoder consisted of three \ac{conv} layers followed by nine residual blocks. The number of (filters, strides) in the first three \ac{conv} layers were set to (32, 1), (64, 2) and (128, 2) respectively. The residual network consisted of two \ac{conv} layers with 128 filters. The decoder network consisted of two de-\ac{conv} layers with strides 2 and filters 64 and 32 respectively followed by a final \ac{conv} layer with stride 1. Instance normalization was used in each layer except in the first and last. ReLU activation was used in all layers except the last. The kernel size in all layers was set to $3\times3$. We used a short cut connection from input \ac{SEN} to the output (input was added to the output of the last layer which becomes \ac{SEN}'s final output). The generators used in \ac{CycleGAN} have similar architecture as the \ac{SEN}.  
The discriminator had 5 \ac{conv} layers each with a kernel size of 4. The strides of first three and last two layers were set to 2 and 1 respectively. The number of filters in each layer were set to 64, 128, 256, 512 and 1. LeakyReLu with slope 0.2 was used as activation in all layers except the last. More details on the architecture can be found in \cite{nidadavolu2019lr}.
 
\subsection{Training Details}
\label{ssec:training_details}
Both the \ac{ENs} and \ac{DAN} were trained on 40-D log \ac{melFB}s extracted from their respective training corpora (details in Sec.\ref{ssec:dataset_details}). The \ac{SEN} was trained to optimize a multi-task objective --combination of \ac{FM} and \ac{adv} losses (details in Sec.~\ref{ssec:sup_enhancement}), with \textit{paired} data drawn from the training corpora. \ac{UEN} and \ac{DAN} were trained to optimize a combination of \ac{cyc} and \ac{adv} losses with \textit{unpaired} data drawn from their respective training corpora (details in Sec.~\ref{ssec:unsup_enhancement} and Sec.~\ref{ssec:dataset_details}). Rest of the training details remain the same for all the networks and are detailed below. Short-time mean centering and energy based \ac{VAD} was applied on the features. Batch size and sequence length were set to 32 and 127 respectively. The models were trained for 50 epochs. Each epoch was set to be complete when one random sample from each of the utterances of \ac{VC} \textit{clean} has appeared once in that epoch. Adam Optimizer was used with momentum $\beta_1=0.5$.
The learning rates for the \ac{ENs} and discriminators were set to 0.0003 and 0.0001 respectively. The learning rates were kept constant for the first 15 epochs and, then, linearly decreased until they reach the minimum learning rate (1e-6). For \ac{SEN}, the \ac{FM} and \ac{adv} loss weights were set to 1.0 and 0.1 respectively. For \ac{UEN} and \ac{DAN}, the \ac{cyc} and \ac{adv} loss weights were set to 2.5 and 1.0 respectively.

We used \ac{ETDNN}~\cite{villalbajhu,nidadavolu2020unsupervised} based x-vector network in the \ac{SVS}. 
More details on the \ac{ETDNN} and the pipeline can be found in~\cite{villalba2019state, garcia2019speaker}. \ac{ETDNN} was trained on 40-D MFCC features using Kaldi\footnote{Data preparation and training scripts can be found at: \url{https://github.com/jsalt2019-diadet/jsalt2019-diadet}}.
During evaluation, output log \ac{melFB} features of \ac{EN}s were converted to MFCCs by applying \ac{DCT} before forward passing through the x-vector network.

\section{Results}
\label{sec:results}

The \ac{BL} \ac{SVS} (evaluated on original features with no enhancement) was termed as \ac{BL}-\ac{SVS}. The \ac{SVS}s evaluated on features mapped using \ac{SEN}, \ac{UEN} and \ac{DAN} were termed as \ac{SEN}-\ac{SVS}, \ac{UEN}-\ac{SVS} and \ac{DAN}-\ac{SVS} respectively. In Sec.~\ref{ssec:sup_vs_unsup_enh}, we present results on a \ac{SVS} trained without data augmentation on simulated test set, which can be treated as system trained on clean and tested on reverberant speech. In this system, enhancement was done during the evaluation stage and was used to tune the \ac{SEN} and compare it with the previously published \ac{UEN}. \ac{DAN} was not used here since we tailored it to work with an augmented x-vector system by training it to map features to \textit{noise} domain.

In Sec.~\ref{ssec:enh_res_svs_with_data_aug}, we compare our enhancement and \ac{DAN} approaches on a more practical scenario--\ac{BL}-\ac{SVS} trained on data augmentation and tested on real datasets acquired from various conditions. In this case, we experimented with training the \ac{SVS} pipeline on enhanced features while also enhancing the evaluation corpora, as suggested in~\cite{nidadavolu2020unsupervised}. 

\vspace{-2mm}

\subsection{Supervised vs unsupervised enhancement}
\label{ssec:sup_vs_unsup_enh}

The \ac{SEN} network was trained using a multi-task objective: a weighted sum of \ac{FM} and \ac{adv} losses (details in Sec.~\ref{ssec:sup_enhancement}).
We first performed an ablation study on the losses by training two \ac{SEN}s on individual losses.
The \ac{SEN}s trained on \ac{FM} loss only and \ac{adv} loss only were termed as \ac{SEN}1 and \ac{SEN}2 respectively.
Results are in Table \ref{tab:sup_vs_unsup_comparison}.
\ac{SEN}2 is trained with only \ac{adv} loss which learns its own loss function.
Using both \ac{SEN}1 and \ac{SEN}2, \ac{SVS}s perform no better than the \ac{BL}-\ac{SVS} and \ac{UEN}-\ac{SVS}.
However, \ac{SEN}2 yielded better \ac{minDCF} (0.535) compared to \ac{SEN}1 (0.626) on \ac{SITW} \textit{reverb}, which justified the usage of \ac{adv} loss. 
\ac{SEN}3 trained using a combination of both the losses performed better than \ac{BL}-\ac{SVS} and \ac{UEN}-\ac{SVS}.
The results suggests that \ac{FM} approach alone is not enough for dereverberation in feature domain.
We further experimented with tuning the \ac{adv} loss weight.
We experimented with {1.0, 0.1 and 0.01}. Setting \ac{adv} loss weight to 0.01 made it insignificant compared to \ac{FM} loss (\ac{SEN}5 had slight improvements over \ac{SEN}1). We obtained better results with the loss weight of 0.1, system we termed as \ac{SEN}4 in Table.\ref{tab:sup_vs_unsup_comparison}. \ac{SEN}4 yielded 20.5\% and 33.5\% percent relative improvements in terms of \ac{minDCF} on \ac{SITW} and \ac{SITW} \textit{reverb} compared to the \ac{BL}-\ac{SVS} (\ac{UEN} yielded 9.1\% and 23\% relative improvements). For the rest of this work, we use \ac{SEN}4 as the \ac{SEN}. 

\begin{table}[htbp]
    \centering
    \resizebox{\columnwidth}{!}{%
    \begin{tabular}{@{}ccccccc@{}}
        \hline
        \textbf{SVS} & \multicolumn{2}{c} {\textbf{Loss Weights}} & 
        \multicolumn{2}{c} {\textbf{SITW}} &
        \multicolumn{2}{c} {\textbf{SITW \textit{reverb}}} \\ 
        system & $\lambda_{FM}$ & $\lambda_{adv}$  & EER & minDCF & EER & minDCF  \\
        \hline              
        \ac{BL} & - & -  & 5.02 & 0.327 & 6.34 & 0.448 \\
        \hline
        \ac{UEN} & - & -  & 4.77 & 0.297 & 5.63 & 0.345 \\
        \hline
        \ac{SEN}1 & 1.0 & 0.0 & 6.39 & 0.462 & 8.87 & 0.626 \\
        \ac{SEN}2 & 0.0 & 1.0 & 8.28 & 0.442 & 9.91 & 0.535 \\
        \ac{SEN}3 & 1.0 & 1.0 & 4.19 & 0.275 & 5.06 & 0.317 \\
        \ac{SEN}4 & 1.0 & 0.1 & \textbf{4.01} & \textbf{0.260} & \textbf{4.63} & \textbf{0.299} \\
        \ac{SEN}5 & 1.0 & 0.01 & 6.28 & 0.462 & 8.72 & 0.612 \\
        \hline
    \end{tabular}%
    }
    \caption {Comparison of \ac{SEN} vs \ac{UEN} on \ac{SVS} evaluated on \ac{SITW} and \ac{SITW} \textit{reverb}. The \ac{SVS} was trained on \ac{VC} \textit{clean} without data augmentation and enhancement was applied during evaluation. \ac{BL} stands for baseline. \ac{UEN} and \ac{SEN} stand for unsupervised and supervised enhancement networks respectively.}
    \label{tab:sup_vs_unsup_comparison}
\end{table}

\vspace{-7mm}

\subsection{Enhancement for \ac{SVS} trained with data augmentation}
\label{ssec:enh_res_svs_with_data_aug}

The \ac{BL}-\ac{SVS} with data augmentation was trained on combination of \ac{VC} \textit{clean}, \ac{VC} \textit{additive} and \ac{VC} \textit{reverb} (details in Sec.~\ref{ssec:dataset_details} and Table.~\ref{tab:datasets_description}).
We experimented with three different training schemes--all modifying training data of x-vector in different ways.
In the first scheme, the entire training data was enhanced using \ac{SEN}, similar to the homogeneous \ac{UEN}-\ac{SVS} pipeline in~\cite{nidadavolu2020unsupervised}.
In the second scheme, training data consisted of dereverberated/enhanced \ac{VC} \textit{reverb} along with unmodified \textit{clean} and \ac{VC} \textit{additive}.
In the third scheme, we trained on the original training data of \ac{BL}-\ac{SVS} along with its enhanced version.
Since the x-vector network in the third case had double the training data compared to the first two scenarios, it was trained for 1.5 epochs compared to 3 epochs in other cases, thus making the systems comparable.
In all schemes, the \ac{PLDA} was trained on x-vectors extracted from enhanced features only (making the \ac{PLDA}s comparable too).
The three schemes were termed as \ac{SVS}1, \ac{SVS}2, and \ac{SVS}3 respectively. In all three cases, the evaluation data was enhanced. All the three training schemes were repeated for \ac{UEN}, \ac{DAN} and \ac{SEN} separately. 

Results are presented in Table~\ref{tab:uen_svs_with_aug_results_ver2}. BabyTrain  benefited from all three feature mapping approaches and all three \ac{SVS}training schemes. \ac{SEN} yielded better results on \ac{SRI} and BabyTrain compared to \ac{UEN} and deteriorated performance on \ac{AMI}. However,
\ac{DAN}-\ac{SVS}3
yielded improvements on all three datasets. 
\ac{DAN}-\ac{SVS}3 yielded relative improvements on \ac{minDCF} of 2.2\%, 6\% and 31.6\% on \ac{AMI}, \ac{SRI} and BabyTrain respectively. 
\ac{SEN}-\ac{SVS}3 yielded 8.6\% and 26.5\% relative improvements on \ac{minDCF} on \ac{SRI} and BabyTrain respectively but deteriorated performance on \ac{AMI}.   
The best relative improvements and the systems that yielded best results on these three datasets were 2.2\% (\ac{DAN}-\ac{SVS}3), 10.7\% (\ac{SEN}-\ac{SVS}2) and 31.6\% (\ac{DAN}-\ac{SVS}3) respectively.
We observed that the x-vector network in homogeneous system \ac{SVS}1 over-fitted  compared to \ac{SVS}2 and \ac{SVS}3 (large gap between training and validation accuracy was observed) because the enhancement removed the noise from the training data. That explained the superior performance of \ac{SVS}3 (or \ac{SVS}2 in some cases) compared to \ac{SVS}1. 
\begin{table}[htbp]
    \centering
    \resizebox{\columnwidth}{!}{%
    \begin{tabular}{@{}cccccccc@{}}
        \hline
        \textbf{Enh.} & \multicolumn{1}{c}{\textbf{SVS}}  & 
        \multicolumn{2}{c} {\textbf{AMI}} &
        \multicolumn{2}{c} {\textbf{SRI}} &
        \multicolumn{2}{c} {\textbf{BABYTRAIN}}\\ 
        Type & Sys. & EER & minDCF & EER & minDCF & EER & minDCF \\
        \hline              
        \ac{BL} &  \ac{SVS} & 18.79 & 0.688 & 14.55 & 0.583 & 11.72 & 0.551\\
        \hline
        \hline
        \multirow{3}{*}{\ac{UEN}} & \ac{SVS}1 & 18.84($\downarrow$) & 0.688($\uparrow$) & 15.29($\downarrow$) & 0.580($\uparrow$) & 9.88($\uparrow$) & 0.391($\uparrow$) \\
         & \ac{SVS}2 &  19.48($\downarrow$) & 0.689($\downarrow$) & 14.69($\downarrow$) & 0.585($\downarrow$) & 11.45($\uparrow$) & 0.474($\uparrow$) \\
         & \ac{SVS}3 &  18.98($\downarrow$) & 0.682($\uparrow$) & 14.02($\uparrow$) & 0.559($\uparrow$) & 13.14($\downarrow$) & 0.441($\uparrow$) \\
        \hline
        \hline
        \multirow{3}{*}{\ac{SEN}} & \ac{SVS}1 & 19.97($\downarrow$) & 0.697($\downarrow$) & 15.16($\downarrow$) & 0.563($\uparrow$) & 9.61($\uparrow$) & 0.428($\uparrow$) \\
         & \ac{SVS}2 & 18.89($\downarrow$) & 0.692($\downarrow$) & 14.50($\uparrow$) & 0.523($\uparrow$) & 8.70($\uparrow$) & 0.402($\uparrow$) \\
         & \ac{SVS}3 & 20.03($\downarrow$) & 0.706($\downarrow$) & 13.13($\uparrow$) & 0.533($\uparrow$) & 10.09($\uparrow$) & 0.405($\uparrow$) \\
        \hline
        \hline
        \multirow{3}{*}{\ac{DAN}} & \ac{SVS}1  & 18.78($\uparrow$) & 0.713($\downarrow$) & 15.18($\downarrow$) & 0.584($\downarrow$) & 12.32($\downarrow$) & 0.488($\uparrow$) \\
         & \ac{SVS}2 & 19.38($\downarrow$) & 0.704($\downarrow$) & 14.40($\uparrow$) & 0.578($\uparrow$) & 11.07($\uparrow$) & 0.523($\uparrow$) \\
         & \ac{SVS}3 & \textbf{18.54($\uparrow$)} & \textbf{0.673($\uparrow$)} & \textbf{14.10($\uparrow$)} & \textbf{0.550($\uparrow$)} & \textbf{8.71($\uparrow$)} & \textbf{0.377($\uparrow$)}  \\
        \hline
        \hline
    \end{tabular}%
    }
    \caption {Results on \ac{SVS} trained with data augmentation ($\uparrow$ and $\downarrow$ indicate that the enhancement system's performance \textit{improved} or \textit{deteriorated} respectively compared to the \ac{BL}-\ac{SVS}. Results in bold indicate system has improved performance across all test conditions. For description of terms \ac{SVS}1, \ac{SVS}2 and \ac{SVS}3 refer to Sec.~\ref{ssec:enh_res_svs_with_data_aug})}
    \label{tab:uen_svs_with_aug_results_ver2}
\end{table}

\vspace{-9mm}

\section{Summary}
\label{sec:summary}

The aim of this study was to make \ac{SVS}s robust to far-field data
using proposed Supervised Enhancement Network (SEN) and Domain Adaptation Network (DAN). 
SEN maps far-field evaluation features to clean domain. It was trained on \textit{paired} data using 
a supervised objective combined with a generative adversarial objective. Meanwhile, 
DAN maps the features to a \textit{noise} domain, which is similar to the augmented data used to train the \ac{SVS} x-vector. \ac{DAN} is trained on \textit{unpaired} data using a CycleGAN scheme. 
We observed that training the \ac{SVS} systems on both original/augmented features and their enhanced version using the networks proposed yielded significant improvements compared to training on augmented data alone or enhanced augmented alone. We observed relative improvements ranging from 2-31\% in terms of minDCF on several simulated and real datasets using this approach. Though the enhancement procedure in this work was targeted at improving \ac{SV} performance, the networks were trained with task independent objectives. Future direction would be to test these techniques on \ac{ASR} task.

\bibliographystyle{IEEEtran}

\bibliography{mybib}


\end{document}